\documentclass[a4paper,11pt]{article}

\title{Einstein gravity as a 3D conformally invariant theory}
\author{\bf Henrique Gomes\footnote{\href{mailto:gomes.ha@gmail.com}{gomes.ha@gmail.com}}\\\it University  of  Nottingham, School of Mathematical Sciences\bigskip\\ \bf Sean Gryb\footnote{\href{mailto:sgryb@perimeterinstitute.ca}{sgryb@perimeterinstitute.ca}} \\\it Perimeter Institute for Theoretical Physics\\\it 31 Caroline Street, Waterloo, Ontario N2L 2Y5, Canada \\ \it and\\ \it University of Waterloo, Department of Physics and Astronomy\\\it Waterloo, Ontario N2L 3G1, Canada \bigskip\\ \bf Tim Koslowski\footnote{\href{mailto:tkoslowski@perimeterinstitute.ca}{tkoslowski@perimeterinstitute.ca}}
\\\it Perimeter Institute for Theoretical Physics\\\it 31 Caroline Street, Waterloo, Ontario N2L 2Y5, Canada}

\usepackage{amssymb}
\usepackage{amsmath}
\usepackage[usenames,dvipsnames]{color}
\usepackage{hyperref}
\usepackage[utf8x]{inputenc}
\usepackage[left=1in,right=1in,top=1in,bottom=1in]{geometry}                  

\hypersetup{
    colorlinks=true,         
    linkcolor=blue,          
    citecolor=red,        
    urlcolor=Violet             
}

\let\oldmarginpar\marginpar
\renewcommand\marginpar[1]{\oldmarginpar{\color{red}\raggedright\scriptsize #1}}

\newcommand{\pb}[2]{\ensuremath{\lf\{#1,#2 \rt\}}}
\newcommand{\mean}[1]{\ensuremath{\lf\langle #1 \rt\rangle }}
\newtheorem{proposition}{Proposition}

\def\lf {\ensuremath{\left}}
\def\rt {\ensuremath{\right}}
\def\absk {\ensuremath{\lf| k \rt|}}
\def\wts {\ensuremath{\widetilde{TS}}}

\begin{document}

\maketitle
\begin{abstract}
    We give an alternative description of the physical content of general relativity that does not require a Lorentz invariant spacetime. Instead, we find that gravity admits a dual description in terms of a theory where local size is irrelevant. The dual theory is invariant under foliation preserving 3--diffeomorphisms and 3D conformal transformations that preserve the 3--volume (for the spatially compact case). Locally, this symmetry is identical to that of Ho\v rava--Lifshitz gravity in the high energy limit but our theory is equivalent to Einstein gravity. Specifically, we find that the solutions of general relativity, in a gauge where the spatial hypersurfaces have constant mean extrinsic curvature, can be mapped to solutions of a particular gauge fixing of the dual theory. Moreover, this duality is not accidental. We provide a general geometric picture for our procedure that allows us to trade foliation invariance for conformal invariance. The dual theory provides a new proposal for the theory space of quantum gravity.
\end{abstract}
\newpage
\tableofcontents

\section{Introduction}

Conformal methods have a long and rich history in the study of gravity. Only two years after Einstein's seminal paper \cite{einstein:gen_rel} on general relativity, Weyl had already proposed a theory \cite{weyl:conformal_1918,weyl:weyl_book}, invariant under conformal transformations of spacetime, as a way to unify gravity with electromagnetism. Unfortunately, no physically sensible theory with this symmetry has been proposed to date.  Attempts to implement 3D conformal invariance have met greater success. As early as 1959, Dirac \cite{Dirac:CMC_fixing} considered gauge fixing the spacetime foliations of general relativity using constraints that generate 3D conformal transformations. These efforts culminated in York's discovery \cite{York:york_method_long,York:york_method_prl} that the initial value problem for general relativity could be solved using conformal methods. Recently, spatial conformal symmetry has played an important role in quantum gravity both through Ho\v rava's proposal \cite{Horava:lif_point} for a UV completion of general relativity and Maldecena's AdS/CFT conjecture \cite{Maldacena:ads_cft}.

In this paper we show that, in gravity, there is a deep relationship between foliation invariance and 3D conformal invariance: they are different ways of expressing the same physical dynamics. The solutions of a particular gauge fixing of general relativity are equivalent to the solutions of a particular gauge fixing of a theory invariant under 3D conformal transformations. In the spatially compact case, which we treat exclusively in this paper, the 3D conformal transformations must preserve the volume of space. This means that there exists an alternative way of understanding gravity that does not require a Lorentz invariant spacetime but relies instead on the relativity of local scale. We will demonstrate that the foliation invariance of general relativity can be traded for 3D local scale invariance and that the mechanism behind the trading has a simple geometric interpretation. Our result is inspired by York's solution to the initial value problem of general relativity.

York began his 1973 paper \cite{York:york_method_prl} on the conformal approach to the initial value problem by stating: ``An increasing amount of evidence shows that the true dynamical degrees of freedom of the gravitational field can be identified directly with the conformally invariant geometry of three--dimensional spacelike hypersurfaces embedded in spacetime.'' He continues: ``the configuration space that emerges is not superspace (the space of Riemannian three--geometries) but `conformal superspace'[the space of which each point is a conformal equivalence class of Riemannian three--geometries]$\times$[the real line](ie, the time, $T$).''(Original parentheses.) Despite these bold claims, he does not show that the dynamics actually projects down to curves on conformal superspace nor does he provide an explanation for how the 4D diffeomorphism invariance of general relativity could be related to 3D conformal invariance. In this paper, we fill in these gaps.

In a recent paper, Barbour and \' O~Murchadha \cite{Barbour:shape_dynamics} make use of an older result \cite{Barbou:CS_plus_V} to show that the dynamical solutions of general relativity on compact manifolds are uniquely determined by initial data on conformal superspace. This is possible provided the spacetime is foliable by spacelike hypersurfaces with constant mean extrinsic curvature (CMC).\footnote{The restriction to CMC foliable spacetimes is weak since it includes the vast majority of physically interesting solutions to Einstein's equations while excluding many physically uninteresting solutions such as those that contain closed timelike curves or those that have exotic topologies.}  They call this theory \emph{shape dynamics} because it involves the evolution of local, scale--independent shapes. The first goal of our paper is to provide a rigorous geometric understanding of this result. We then extend this result by uncovering the precise relationship between general relativity and its dual theory -- presumably Barbour and \'O~Murchadha's shape dynamics.

We establish our result by introducing a St\"uckelberg field\footnote{We use the term St\"uckelberg field, because the mechanism is precisely the St\"uckelberg's mechanism of restoring U(1)-gauge symmetry in massive electrodynamics \cite{stueckelberg}.} that is subsequently eliminated from the theory after performing a particular gauge fixing. This process and, in particular, the gauge fixing is inspired by the ``best matching'' used in \cite{Barbou:CS_plus_V}. For a detailed account of this procedure, that was first developed in \cite{barbourbertotti:mach}, see \cite{barbour:bm_review} (or \cite{sg:dirac_algebra} for a canonical treatment).

Our results both clarify and justify York's intuitive remarks regarding the configuration space of gravity: conformal superspace is not the reduced configuration space of general relativity but that of its dual theory. This fact could have important implications for quantum gravity. In the dual theory that we will develop, there is one global (non--local) Hamiltonian that generates the dynamics of the system on conformal superspace. The dual theory then suggests a natural candidate for the \emph{physical} Hilbert space of quantum gravity: the set of states invariant under foliation preserving diffeomorphisms and volume preserving\footnote{They need only be volume preserving in the spatially compact case where the volume is well defined.} conformal transformations (this also implies a similarly defined theory space). Given the many difficulties encountered in defining the physical Hilbert space of quantum general relativity, it may be useful to define quantum gravity as the quantization of the dynamics of shape.

\subsection{Outline}

The outline of the paper is as follows: we first illustrate our procedure for trading symmetries by applying it to a simple toy model. We then develop the general geometric picture that justifies our algorithm. Finally, we apply this algorithm to geometrodynamics and establish a dictionary that maps solutions of the dual theory to solutions of general relativity. In an appendix, we provide further mathematical details supporting the result on gravity.

The toy model is a finite dimensional particle model constructed to exhibit many of the features of geometrodynamics. We've included it both for pedagogical purposes and because it provides an explicit example where all the equations can be worked out and the symmetry trading can be verified directly. However, this section can be skipped on a first reading if the reader is interested only in the main result.

\section{Duality in Toy Particle Model}

The toy model consists of $n$ independent harmonic oscillators in $d$ dimensions. The action is\footnote{Lowercase indices are spatial ranging from $1$ to $d$ and repeated indices are summed over. Uppercase indices label particle number and are only summed when explicitly shown. Also, $v^2 \equiv v^a v^b \delta_{ab}$ for vectors and $u^2 \equiv u_a u_b \delta^{ab}$ for covectors.}
\begin{equation}
    S = \int_{t_0}^{t_1} dt\, \sum_{I=1}^{n} \frac 1 2 \left[ \lf(\frac{dq_I}{dt}(t) \rt)^2  - k q^2_I(t) \right].
\end{equation}
The masses have been set to 1, for convenience. We will need to consider negative spring constants $k<0$ to avoid imaginary solutions. Unstable solutions will not worry us here since we are interested mainly in the symmetries of the toy model, which we do not consider a physical theory.

We write the above action in a reparametrization invariant form by introducing the auxiliary fields $N_I$ and the arbitrary label $\lambda$
\begin{equation}
    S = \int^{\lambda_1}_{\lambda_0} d\lambda\, \sum_{I=1}^{n} \left[ \frac{\dot q^2_I(\lambda)}{2N_I}  + N_I \lf( \frac{\absk}{2} q^2_I(\lambda) + \mathcal{E}_I \rt) \right],
\end{equation}
where we have restricted to strictly negative $k$'s. We choose to parametrize the trajectory of each particle independently and allow the time variables to be varied dynamically. The $N_I$'s represent the $\lambda$ derivatives of the time variables and mimic the local lapse functions of geometrodynamics, whose terminology we will borrow. The total energy of each particle $\mathcal{E}_I$ essentially specifies how much time should elapse for each particle in the interval $\lambda_1 - \lambda_0$ so that this theory is equivalent to $n$ harmonic oscillators with different ``local'' time variables. We have introduced the local reparametrization invariance as a way of studying this feature of geometrodynamics in a finite dimensional toy model. It is not meant to have any direct physical significance.

After performing the Legendre transform $S = \int d\lambda \sum_{I=1}^n \lf( \pi^I_a \dot q^a_I - H_0(q, \pi)\rt)$, where $\pi^I_a = \frac{\delta S}{\delta \dot q^a_I}$, the local reparametrization invariance leads to $n$ first class constraints, $\chi_I$. The total Hamiltonian is
\begin{align}\label{original hamiltonian}
    H_0 &= \sum_{I=1}^n N^I \chi_I, \text{ where} \\
    \chi_I &= \frac 1 2 \lf( \pi^2_I - \absk q^2_I \rt) - \mathcal{E}_I
\end{align}
and the phase space $\Gamma(q,\pi)$ is equipped with the Poisson bracket
\begin{equation}
    \pb{q^a_I}{\pi_{b}^J} = \delta^a_b\delta_{I}^J.
\end{equation}
Our goal, as explained below, will be to exchange all but one of the first class $\chi_I$'s for new first class conformal constraints.

\subsection{Identify the Symmetries to Exchange}

Our stated objective is to exchange the local reparametrization invariance for conformal invariance. A na\" ive choice would be to exchange all the $\chi_I$'s for the symmetry, $\mathcal S$, defined by identifying the configurations
\begin{equation}
    q_I^a \to e^{\phi_I} q^a_I
\end{equation}
as members of an equivalence class for some arbitrary scale parameters $\phi_I$. This choice, however, will lead to frozen dynamics because one global $\chi$ must be left over to generate dynamical trajectories on phase space\footnote{We draw the reader's attention to \cite{barbour_foster:dirac_thm}, which suggests that such a global $\chi$ does not generate a gauge transformation but a genuine physical
evolution.}. Thus, we need to leave at least one linear combination of $\chi$'s unchanged. Because of this, our first--and most subjective--step is to identify a subset of $\mathcal S$ to exchange.

A natural choice is the symmetry $\mathcal S / \mathcal V$, where $\mathcal V$ represents all configurations that share the same total moment of inertia. Since the total moment of inertia sets the global scale of the system, $\mathcal S / \mathcal V$ identifies all locally rescaled  $q$'s that share the same global scale. This symmetry requires invariance under the transformation
\begin{equation}\label{eq: q symmetry}
    q_I^a \to e^{\hat\phi_I} q^a_I,
\end{equation}
where $\hat\phi$ obeys the identity
\begin{equation}\label{eq: q identity}
    \mean{\hat\phi} = 0.
\end{equation}
The \emph{mean} operator, is defined as $\mean{\cdot} \equiv \frac 1 n \sum_{I=1}^n \cdot_I$, where $\cdot$ signifies inclusion of the field we would like to take the mean of. We can impose the identity (\ref{eq: q identity}) explicitly by writing $\hat\phi$ in terms of the local scale factors $\phi_I$
\begin{equation}\label{eq: hat phi def}
    \hat\phi_I = \phi_I - \mean{\phi}.
\end{equation}
Thus, we parametrize the group $\mathcal S / \mathcal V$ redundantly by $\mathcal S$.

\subsection{Introduce St\" uckelberg Field}

The next step is to artificially introduce the symmetry $\mathcal S / \mathcal V$ by applying the canonical transformation $t$
\begin{align}
    q^a_I &\to t_\phi q^a_I \equiv Q^a_I = e^{\hat\phi_I} q^a_I \notag \\
    \pi^I_a &\to t_\phi\pi^I_a \equiv \Pi^I_a = e^{-\hat\phi_I} \pi_a^I
\end{align}
generated by
\begin{equation}\label{eq: can trans q}
    F(\hat\phi) = \sum_I q^a_I \exp(\hat\phi_I) \Pi^I_a.
\end{equation}
We then enlarge the phase space $\Gamma(q,\pi) \to \Gamma_\text{e}(q, \pi, \phi, \pi_\phi)$ to include the canonically conjugate variables $\phi_I$ and $\pi_\phi^I$ that we use to construct $\hat\phi$ using (\ref{eq: hat phi def}). The non--zero Poisson bracket's introduced in $\Gamma_\text{e}$ are
\begin{equation}
    \pb{\phi_I}{\pi_\phi^J} = \delta^J_I.
\end{equation}
We can naturally extend $t$ to a map $t:\Gamma_\text{e}\rightarrow\Gamma_\text{e}$ by acting in the following way: $t:(q,\pi,\phi,\pi_\phi)\mapsto (t_\phi q,t_\phi \pi,\phi,\pi_\phi)$. This gives the natural extension of functions, $f\in C^\infty(\Gamma)$, on $\Gamma$ to functions, $\tilde f=f\circ p\circ t\in C^\infty(\Gamma_\text{e})$,  on $\Gamma_\text{e}$ induced by the canonical transformation (\ref{eq: can trans q}):
$$\tilde f(q,\pi,\phi,\pi_\phi)=f(t_\phi q,t_\phi \pi).$$
Here, $p:\Gamma_\text{e}:=\Gamma\times T^*\mathcal{S}\rightarrow \Gamma$ is a canonical projection. From now on, we will abbreviate the transformation\footnote{This transformation is analogous to the restoration of $U(1)$-gauge symmetry in massive electrodynamics through the indroduction of a St\"uckelberg field $\phi$: The original photon field $A_\mu$ is replaced by a $U(1)$-transformed $A_\mu+\partial_\mu \phi$, where $\phi$ is not viewed as a gauge parameter but as a new field with its own gauge transformation property, such that the transformed photon mass term $\frac{m^2}2 (A^\mu+\partial^\mu \phi) (A_\mu+\partial_\mu \phi)$ is gauge-invariant.} above $(p\circ t)^*:=T$.

The Poisson structure of $\Gamma_\text{e}$ is such that
\begin{equation} \label{eq: poisson relation}
    \pb{Tf(q,\pi)}{\pi_\phi^I} = \pb{Tf(q,\pi)}{\pi^I_a q^a_I - \mean{\pi\cdot q}},
\end{equation}
In the above, we have used the fact that $\pi^I_a q^a_I - \mean{\pi\cdot q}$ is invariant under $T$, as can be checked explicitly. Consistency between the dynamics and this Poisson structure requires the constraints\footnote{These constraints could have been derived as primary constraints by introducing $\hat\phi$ \emph{before} the Legendre transform through ``best matching''. See \cite{barbour:bm_review} or \cite{sg:dirac_algebra} for details.}
\begin{equation}
    C^I = \pi_\phi^I - \lf( \pi^I_a q^a_I - \mean{\pi\cdot q} \rt)\approx 0.
\end{equation}

Given (\ref{eq: poisson relation}), the $C^I$'s are trivially first class with respect to the transformed constraints $T\chi_I$
\begin{equation}
    \chi_I \to T\chi_I = \frac 1 2 \lf( e^{-2\hat\phi} \pi_I^2 - e^{2\hat\phi} \absk q_I^2 \rt)
\end{equation}
as they commute with \emph{any} $f(Q,\Pi)$. This should be expected: the canonical transformation (\ref{eq: can trans q}) did not change the theory. Since they act trivially on the functions extended into $\Gamma_e$, the original theory is unmodified. But, by extending the phase space, we have added two phase space degrees of freedom per particle. The role of the $C_I$'s is precisely to remove these degrees of freedom.  The $\hat\phi$ field is essentially a Stueckelberg field for the conformal symmetry. This can only happen if the $C_I$'s are first class. Consequently, the infinitesimal gauge transformations generated by the $C_I$'s on the extended phase space variables
\begin{align}
    \delta_{\theta_J} q_I &= -\delta^{IJ}\hat\theta_I q_I & \delta_{\theta_J} \pi_I &= \delta^{IJ}\hat\theta_I\pi_I \nonumber\\
   \label{eq: C_I extended trans} \delta_{\theta_J} \phi_I &= \delta^{IJ}\theta_J & \delta_{\theta_J} \pi^I_\phi &= 0,
\end{align}
where $\delta_{\theta_J} \cdot = \pb{\cdot}{ \theta^J C_J}$ (no summation), represents at this stage a completely artificial symmetry.

Before writing down the remaining constraints, we note that one of the $C_I$'s is singled out by our construction. Averaging all the $C_I$'s gives
\begin{equation}\label{eq: pi identity}
    \mean{\pi_\phi} \approx 0.
\end{equation}

However, the action of this constraint is trivial for arbitrary functions $f$ on the extended phase space $\Gamma_\text{e}$. To see this, note that the gauge transformations $\delta_\theta \cdot = \pb{\cdot}{\theta\mean{\pi_\phi}}$ generated by (\ref{eq: pi identity}) are trivial for the phase space variables $q$, $\pi$, $\hat\phi$, and $\pi_\phi$. The only transformation that is not automatically zero is $\delta_\theta \hat\phi_I = \theta \pb{\hat\phi_I}{\mean{\pi_\phi}}$, which can easily be seen to vanish. Since the $\phi$'s only enter the theory through the $\hat\phi$'s, we see that the constraints (\ref{eq: pi identity}) act trivially on all quantities in our theory. This is a reflection of the fact that there is a redundancy in the $\hat\phi$ variables. Eq~(\ref{eq: pi identity}) is then an identity and not an additional constraint. This reflects the redundancy in the original $\hat\phi$ variables expressed in terms of $\phi$'s.

It will be convenient to separate this identity from the remaining constraints by introducing the numbers $\alpha^I_i$, where $i = 1\dots (n-1)$, such that the set $\lf\{\mean{\pi_\phi}, C_i \rt\}$, where $C_i = \sum_I \alpha^I_i C_I$, forms a linearly independent basis for the $C_I$'s.\footnote{Note that the ability to explicitly construct this basis is a feature of the finite dimensional model. This possibility sets the toy model apart from the situation in field theory, which is more subtle.} We are then safe to drop the constraint $\mean{\pi_\phi} = 0$ from the theory. The remaining constraints are now
\begin{align}
    T\chi_I &\approx 0 , \text{   and} & C_i \approx 0.
\end{align}
where $I = 1,\dots,n$ and $i = 1,\dots,n-1$.

\subsection{Impose and Propagate Best--Matching Constraints}\label{sec: Mach conditions}

In order to exchange symmetries, we impose the \emph{best--matching} (also known as \emph{Mach} constraints) constraints
\begin{equation}
    \pi_\phi^I \approx 0.
\end{equation}
These constraints are motivated by the desire make the theory background independent with respect to the $\mathcal C/\mathcal V$ symmetry. For more details on how these constraints achieve this see, \cite{barbour:scale_inv_particles,sg:dirac_algebra}. One of the best--matching constraints, $\mean{\pi_\phi} \approx 0$, is first class but trivially satisfied. It can be dropped. The remaining $\pi_\phi^I$'s can be represented in a particular basis as $\pi_\phi^i = \alpha^i_I \pi_\phi^I$ and are second class with respect to the $T\chi$'s. To see this, consider the smearing functions $f^I$. Then, the Poisson bracket
\begin{equation}
    \sum_{I,J=1}^{n} \pb{f^IT\chi_I}{ \alpha^j_J\pi^J_\phi} = \sum_{J=1}^n \lf[ \mean{\alpha^j_J f^J \lf(  \pi^2_J + \absk q^2_J \rt)} - \alpha^j_J f^J \lf( \pi^2_J + \absk q^2_J \rt) \rt]
\end{equation}
is non--zero for general $f^I$ (in the above, the mean is taken with respect to the $j$ components). Nevertheless, the constraints $\pi_\phi^i$ can be propagated by the Hamiltonian by using the preferred smearing $f^I = N^I_0$ that solves the $n-1$ equations
\begin{equation} \label{eq: lf particle}
    \sum_{J=1}^n \alpha^j_J N_0^J \lf( \absk q^2_J + \pi^2_J \rt) = \mean{\alpha^j_J N_0^J \lf( \absk q^2_J + \pi^2_J \rt)}.
\end{equation}
These lapse fixing equations lead to consistent equations of motion.

We can now understand the reason for imposing the identity $\mean{\hat\phi} = 0$. Were $\hat\phi$ unrestricted, the $n\times n$ matrix $\pb{T\chi_I}{\pi_\phi^J}$ would have a trivial kernel $N^I_0 = 0$. This uninteresting solution leads to frozen dynamics. Alternatively, the $(n-1)\times n$ matrix $\pb{T\chi_I}{\pi_\phi^i}$ has a one dimensional kernel. Thus, (\ref{eq: lf particle}) has a one parameter family of solutions parameterized by the global lapse. This leaves us, in the end, with a non--trivial time evolution for the system.

The fact that we have a remaining global lapse indicates that we have a remaining linear combination of $T\chi_I$'s that is first class. This is the linear combination given by the preferred smearing $N_0^I$. We can split the $T\chi_I$'s into a single first class constraint
\begin{equation}
    T\chi_\text{f.c.} = \sum_I N_0^I T\chi_I
\end{equation}
and the second class constraints
\begin{equation}
    T\chi_i = \sum_I \beta_i^I T\chi_I,
\end{equation}
where the $\beta_i^I$ are numbers chosen so that the above set of constraints is linearly independent. Note that, in the finite dimensional case, it is easy to ensure that this new basis is equivalent to the original set of $\chi_I$'s. In geometrodynamics, the situation is more subtle because we are dealing with continuous degrees of freedom. Using this splitting, we now have the first class constraints
\begin{align}\label{eq 1st class part}
    T\chi_\text{f.c.} &\approx 0, \text{   and} & C_i &\approx 0
\end{align}
as well as the second class constraints
\begin{align}\label{eq 2nd class part}
    T\chi_i &\approx 0, \text{   and} & \pi^i_\phi &\approx 0.
\end{align}

\subsubsection{Excursion: Best Matching}

Since best--matching is just an inspiration for the construction in our paper, we restrict our explanation of best-matching to the minimum. 

In the construction of an action principle one is often faced with the problem that one can not directly parametrize the physical configuration space of a system. However, one can often describe the configurations space as the quotient of a redundant parametrizable space by the action of a gauge group, that one can parametrize as well. A prime example for this is geometrodynamics, where the distinct configurations, i.e. points in superspace, are not easily accessible, but can be described as diffeomorphism orbits in the space Riem, which consists of all 3-metrics. Each of these diffeomorphism orbits is a subspace of Riem. If we want to construct a geometrodynamical action, we have to construct it as an action for these subspaces.

Let us now for simplicity assume that the action principle is given by the geodesic length of a curve in the redundant configuration space, for some given metric, and that the physical configurations are represented by non-intersecting subspaces of the redundant configuration space. The action principle asks for the minimal distance between two subspaces, so we look for the shortest (best--matching) geodesic that starts at one and ends at the other. The action thus gives two conditions (1) the curve has to be a geodesic and (2) it has to intersect both subspaces orthogonally. These two conditions generalize to (1) the curve has to extremize the action and (2) the conjugate momenta have to be orthogonal\footnote{As a one-form, it in fact has to have the submanifold as its nullifier.}. One can straightforwardly see that if $\gamma$ parametrizes the gauge group (and hence the orbit manifold) that this condition on the conjugate momenta is equivalent to demanding that the best--matching condition
$$ 
  \pi_\gamma=\frac{\partial L}{\partial \dot\gamma}=0
$$
holds. One can furthermore straightforwardly see that for the 3-diffeomorphism group the condition one arrives at is the momentum constraint. For more details on this derivation, see \cite{Gomes:gauge_theory_riem}.

\subsection{Eliminate Second Class Constraints}

To see that this procedure has succeeded in exchanging symmetries, we can eliminate the second class constraints by defining the Dirac bracket
\begin{equation}
    \pb{\cdot}{\cdot\cdot}_\text{Db} = \pb{\cdot}{\cdot\cdot}_\text{Pb} + \sum_{i,j = 1}^{n-1} \pb{\cdot}{\pi_\phi^i} C_i^j \pb{T\chi_j}{\cdot\cdot} - \pb{\cdot}{T\chi_i} C^i_j \pb{\pi_\phi^j}{\cdot\cdot},
\end{equation}
where
\begin{equation}
     C^i_j = \pb{\pi_\phi^i}{T\chi_j}^{-1}.
\end{equation}
The existence and uniqueness of the Dirac bracket depend on the invertibility of $C$, which, in turn, depends on the existence and uniqueness of the lapse fixing equations (\ref{eq: lf particle}). In this toy model, (\ref{eq: lf particle}) are just arbitrary linear algebraic equations. Thus, $C$ will generally have a non--vanishing determinant.

The advantage of using the Dirac bracket is that the second class constraints $\pi_\phi^i$ and $T\chi_i$ become first class and can be applied \emph{strongly} and eliminated from the theory. This procedure is greatly simplified because one of our second class constraints, namely $\pi_\phi^i = 0$, is proportional to a phase space variable. Thus, the method discussed by Dirac in \cite{Dirac:CMC_fixing} can be applied to eliminate the second class constraints. For $\pi_\phi^i$, this involves setting $\pi_\phi^i = 0$ everywhere in the action. For the $T\chi_i$'s, this involves treating $T\chi_i$ as an equation for $\phi$. The solution, $\phi_0(q, \pi)$, of
\begin{equation}\label{eq phi_0 particle}
    \lf. T\chi_i\rt|_{\phi^I = \phi^I_0} =0
\end{equation}
is then inserted back into the Hamiltonian. This leaves us with the dual Hamiltonian
\begin{equation}
    H_\text{dual} = N T\chi_\text{f.c.}(\phi_0) + \sum_{i=1}^{n-1} \Lambda^i \sum_{I=1}^n \alpha^i_I\lf( q^a_I \pi^I_a - \mean{q \cdot \pi} \rt),
\end{equation}
where $N$ and $\Lambda^i$ are Lagrange multipliers. We can write this in a more convenient form by using the redundant constraints $D_I = q^a_I \pi^I_a - \mean{q \cdot \pi}$
\begin{equation} \label{eq dual ham particle}
    H_\text{dual} = N T\chi_\text{f.c.}(\phi_0) + \sum_{I=1}^{n} \Lambda^I D_I
\end{equation}
remembering that one of the $D_I$'s is trivially satisfied.

The remaining Hamiltonian is now dependent only on functions of the original phase space $\Gamma$. Furthermore, the Dirac bracket between the dual Hamiltonian and any functions $f(q,\pi)$ on $\Gamma$ reduces to the standard Poisson bracket. This can be seen by noting that the extra terms in the Dirac bracket $\pb{f(q,\pi)}{H_\text{dual}}_\text{Db}$ only contain a piece proportional to $\pb{f(q,\pi)}{\pi_\phi}$, which is zero, and a piece proportional to, $\pb{\pi_\phi^i}{H_\text{dual}}$, which is weakly zero because $H_\text{dual}$ is first class. Thus, this standard procedure eliminates $\phi$ and $\pi_\phi$ from the theory.

The theory defined by the dual Hamiltonian (\ref{eq dual ham particle}) has the required symmetries: it is invariant under global reparametrizations generated by $\lf. \chi_\text{f.c.}\rt|_{\phi_0}$ and it is invariant under scale transformations that preserve the moment of inertia of the system. This last invariance can be seen by noting that the first class constraints $D_I$ generate exactly the symmetry (\ref{eq: q symmetry}).

Before leaving this section we will give two explicit examples of how one can construct $T\chi_\text{f.c.}(\phi_0)$ explicitly. In the first example, we solve the second class $T\chi_i$'s for $\phi_0$ exactly and then plug the solution into the definition of $T\chi_\text{f.c.}$. This immediately gives the general solution for $n$ particles. Unfortunately, this procedure cannot be implemented explicitly in geometrodynamics because the analogue of $T\chi_i(\phi_0) = 0$ is a differential equation for $\phi_0$. There is, however, a slightly simpler procedure that one can follow for constructing $T\chi_\text{f.c.}(\phi_0)$ provided one can find initial data that satisfy the initial value constraints $\chi_i$ of the original theory. If such initial data can be found, then $\phi_0 = 0$ is a solution for these initial conditions. Because the theory is consistent, the initial value constraints will be propagated by the equations of motion. Thus, $\phi_0 = 0$ for all time. We can then use this special solution and compute $\chi_\text{f.c.}$ using its definition. This alternative method is given in the second example for the $n=3$ case and agrees with the more general approach.

\subsubsection{General $n$}

For simplicity, we take $\mathcal E_I = 0$. This will lead to transparent equations, which we value more in the toy model than physical relevance. In the particle model, it is possible to find $\phi^I_0$ explicitly for an arbitrary number of particles, $n$, by solving $T\chi_I(\phi_0) = 0$. It is easiest to solve for $\hat\phi^I_0$ directly. The solution is\footnote{Negative spring constants $k<0$ are required to give real solutions to this equation.}
\begin{equation}\label{eq phi_0}
    e^{2\hat\phi^I_0} = \sqrt{\frac{\pi^2_I}{\absk q^2_I}}.
\end{equation}
We can choose a simple basis for the second class $\chi_I$'s: $\{ \chi_i| i \neq j\}$, for some arbitrary $j$. The first class Hamiltonian is then
\begin{equation}
    \chi_\text{f.c.} = N^j_0 \chi_j(\hat\phi^I_0).
\end{equation}
Using the identity $\hat\phi^j = - \sum_{I\neq j} \hat\phi^I$, we can rewrite this as
\begin{equation}
    \chi_\text{f.c.} \propto \exp\lf( 2\sum_{I\neq j} \hat\phi^I_0 \rt) \pi^2_j + \exp\lf( -2\sum_{I\neq j} \hat\phi^I_0 \rt) \absk q^2_j.
\end{equation}
Inserting (\ref{eq phi_0}), the first class $\chi$ becomes
\begin{equation}\label{eq chi_fc n particle}
    \chi_\text{f.c.} \propto \prod_{I=1}^n \pi^2_I  - \absk^{n} \prod_{I=1}^n q^2_I.
\end{equation}
It can be readily checked that (\ref{eq chi_fc n particle}) is invariant under the symmetry
\begin{align}
    q_I &\to e^{\hat\phi_I} q_I & \pi_I &\to e^{-\hat\phi_I} \pi_I,
\end{align}
where, $\hat\phi^I = \phi^I - \mean{\phi}$.

In geometrodynamics, solving for $\hat\phi$ explicitly will not be possible because it will require the inversion of a partial differential equation. However, if one is only interested in mapping solutions from one side of the duality to the other, then it is sufficient to find initial data that satisfy the $\chi_I$'s with $\phi = 0$ and use the preferred lapse to construct the global Hamiltonian. This procedure is outlined in the next example.

\subsubsection{Example: $n = 3$}

We analyse the 3 particle case because this involves subtleties that do not arise in the 2 particle case but must be dealt with in the general $n$ particle case.

If we choose the basis $\{\pi^i_\phi\} = \{\pi^1_\phi, \pi^2_\phi\}$, then the lapse fixing equations (\ref{eq: lf particle}) with $\phi = 0$ have the unique solution
\begin{align}
    N^1_0 &= N^3_0 \lf( \frac{\absk q^2_3 + \pi^2_3}{\absk q^2_1 + \pi^2_1} \rt) &  N^2_0 &= N^3_0 \lf( \frac{\absk q^2_3 + \pi^2_3}{\absk q^2_2 + \pi^2_2} \rt).
\end{align}
Inserting this solution into $\lf. \chi_\text{f.c.}\rt|_{\phi = 0} =\sum_I N^I_0 \chi_I$, we find
\begin{equation}
    \chi_\text{f.c} \propto \lf( \pi^2_1 \pi^2_2 \pi^2_3 - \absk^3 q^2_1 q^2_2 q^2_3  \rt) - 2\chi_1 \chi_2 \chi_3.
\end{equation}
The consequence of setting $\phi^I = 0$ instead of using it to solve $T\chi_I(\phi_0^I) = 0$ is that we pick up a term that is proportional to $\chi_1$ and $\chi_2$, which are \emph{strongly} zero. For this reason, we need to ensure that we have initial data that solve the initial value constraints ($\chi_1 = 0$ and $\chi_2 = 0$ in this case) for $\phi = 0$ in order to arrive at the correct global Hamiltonian. Assuming that such data have been found, the last term is zero and $\chi_\text{f.c.}$ reduces to
\begin{equation}\label{eq chi_fc 3 particle}
    \chi_\text{f.c} \propto \lf( \pi^2_1 \pi^2_2 \pi^2_3 - \absk^3 q^2_1 q^2_2 q^2_3 \rt).
\end{equation}
This agrees with our general result from last section.

\subsection{Construct Explicit Dictionary}

We can now compare our starting Hamiltonian with that of the dual theory:
\begin{align}
    H_0 &= \sum_I N^I \chi_I \\
    H_\text{dual} &= N\lf. \chi_\text{f.c.}\rt|_{\phi_0} + \sum_I \Lambda^I \lf( \pi^I_a q^a_I - \mean{ \pi_I \cdot q_I} \rt) \notag \\
                  &= N\lf[ \prod_{I=1}^n \pi^2_I  - \absk^{n} \prod_{I=1}^n q^2_I \rt] + \sum_I \Lambda^I \lf( \pi^I_a q^a_I - \mean{ \pi_I \cdot q_I} \rt).
\end{align}
The first theory is locally reparametrization invariant while the second has local scale invariance and is only globally reparametrization invariant. Note the highly non--local nature of the global Hamiltonian of the dual theory.

The equations of motion of the original theory are
\begin{align}
    \dot q^a_I &= N^I \pi^I_b \delta^{ab} & \dot \pi^I_a &= N^I \absk q^b_I \delta_{ab}.
\end{align}
Those of the dual theory are
\begin{align}
    \dot q^a_I &= \lf[2N \prod_{J\neq I} \pi^2_J \rt] \pi^I_b \delta^{ab} + \frac{n-1}{n}\Lambda_I q^a_I \\
    \dot \pi^I_a &= \lf[2N \prod_{J\neq I} \absk q^2_J \rt] \absk q^b_I \delta_{ab} + \frac{1-n}{n}\Lambda_I \pi^I_a.
\end{align}
We can now read off the choice of Lagrange multipliers for which the equations of motion are equivalent
\begin{align}\label{eq N dic}
    N^I &= 2N \prod_{J\neq I} \pi^2_J \approx 2N \prod_{J\neq I} \absk q^2_J \\
    \Lambda^I &= 0.
\end{align}
In the second equality of (\ref{eq N dic}), the $\chi_I$'s have been used.

\section{General Symmetry Trading Algorithm}

Based on the above description of the procedure, we can sketch out a general algorithm for trading the first class symmetry $\chi$ for $D$:
\begin{enumerate}
    \item Artificially introduce the symmetry to be gained in the exchange by performing a canonical transformation of the schematic form:
    \begin{equation}
	F(\phi) = q \exp(\phi) \Pi.
    \end{equation}
	This extends the phase space of the theory to include the Stueckelberg field $\phi$ and its conjugate momentum $\pi_\phi$ and introduces constraints of the form
    \begin{equation}
	    C = D - \pi_\phi.
    \end{equation}
    \item Impose the best--matching conditions
	\begin{equation}
	    \pi_\phi \approx 0.
	\end{equation}
    If they are first class, then the procedure just ``gauges'' a global symmetry. If they are second class with respect to the transformed first class constraints $T\chi$ of the original theory, then they can be exploited for the exchange if the Lagrange multipliers, $N^0$, can be fixed uniquely such that
    \begin{equation}
	\pb{T\chi(N^0)}{\pi_\phi} = 0.
    \end{equation}
    When this is possible, the $\pi_\phi$'s can be treated as special gauge fixing conditions for $\chi$. It is possible, as in our case, that there will be a part of $T\chi$ that is still first class with respect to $\pi_\phi$. This should be separated from the purely second class part.
    \item Define the Dirac bracket to eliminate the second class constraints. This can be done provided the operator $\pb{T\chi}{\pi_\phi}$ has an inverse (which also means that $N^0$ exists and is unique). The second class constraints can be solved by setting $\pi_\phi = 0$ everywhere in the Hamiltonian and by setting $\phi = \phi_0$ such that $T\chi(\phi_0) = 0$. The second class condition and the implicit function theorem guarantee that this can be done. When this is done, the Dirac bracket reduces to the Poisson bracket on the remaining phase space variables. The $\pi_\phi$ terms drop out of the $C$'s and the $\chi$'s have been traded for $D$'s.
    \item To check consistency, construct the explicit dictionary by reading off the gauges in both theories that lead to equivalent equations of motion.
\end{enumerate}

Remark: We first pass from an original gauge theory on phase space $\Gamma$ to a theory on a larger phase space $\Gamma_{e}$, containing the St\"uckelberg field $\phi$ as a new physical field and not a gauge parameter. The new theory is however precisely the original if one imposes the $C$ constraints and chooses a gauge such that $\phi=0$.

Guided by this basic algorithm, we will now construct a formal geometric picture to illustrate why this algorithm works. 

\subsection{Geometric Picture}

Let us now examine the mathematical structure behind the trading of gauge symmetries that we encountered by constructing the dual theory. We start out with a gauge theory $(\Gamma,\{.,.\},H,\{\chi_i\}_{i\in\mathcal I})$, where $\Gamma$ is the phase space of the theory supporting the Poisson-bracket $\{.,.\}$, the Hamiltonian $H$, and the constraints $\chi_i$. We demand that the constraints are first class, i.e. for all $i,j\in \mathcal I$ there exist phase space functions $f_{ij}^k$ as well as $u_i^k$ s.t.
\begin{equation}\label{equ:first-class-constraints}
 \begin{array}{rcl}
   \{\chi_i,\chi_j\}&=&\sum_{k\in\mathcal I}f^k_{ij} \chi_k\\
   \{H,\chi_i\}&=&\sum_{k\in\mathcal I} u^k_i \chi_k.
 \end{array}
\end{equation}
The constraints define a subspace\footnote{Although we assert ``space'' we abstain from topologizing $\Gamma$ or any of its subsets in this paper.} $\mathcal C = \{x\in \Gamma: \chi_i(x)=0\,\, \forall i\in\mathcal I\}$ and the first line of equation \eqref{equ:first-class-constraints} implies that the derivations $\{\chi_i,.\}$ are tangent to $\mathcal C$, while the second line implies that the Hamilton vector field $\{H,.\}$ is tangent to $\mathcal C$. The $\{\chi_i,.\}$ generate a group $\mathbb G$ of gauge-transformations on $\mathcal C$ that lets us identify $\mathcal C$ through an isomorphism $i:E\to\mathcal C$ with a bundle\footnote{This bundle is, in general, not a fibre bundle, since different points $x$ in $\mathcal C$ generally have different isotropy groups $Iso(x)$.} $E$ over $\mathcal C/\mathbb G$. The fibres of $E$ are gauge orbits of a point $x\in \mathcal C$ and thus isomorphic to $\mathbb G/Iso(x)$. According to Dirac, we identify the fibres with physical states and observe that the total Hamiltonian $H_{tot}=H+\sum_{i\in\mathcal I}\lambda^i \chi_i$ depends on a set of undetermined Lagrange multipliers $\{\lambda_i\}_{i\in\mathcal I}$. Fixing a gauge means to find a section $\sigma$ in $E$.

This is most easily done by imposing a set of gauge-fixing conditions\footnote{In general the index set for $\phi_i$ could be different from $\mathcal I$.} $\{\phi_i\}_{i \in \mathcal I}$ such that the intersection of $\mathcal G=\{x\in\Gamma:\phi_i(x)=0\,\,\forall i\in \mathcal I\}$ with $\mathcal C$ coincides with $i(\sigma)$. To preserve $i(\sigma)$ under time evolution we have to solve $\left.\{H_{tot},\phi_i\}\right|_{\mathcal C}=0\,\,\forall i\in\mathcal I$ for the Lagrange-multipliers $\lambda^j=\lambda^j_o$. The gauge-fixed Hamiltonian is
\begin{equation}H_{gf}:=H+\sum_{i\in\mathcal I}\lambda_o^i \chi_j\end{equation} and the equations of motion are generated by $\{H_{gf},.\}$, while the initial value problem is to find data on $i(\sigma)$.\medskip

Given a gauge theory $(H,\{\chi_i\}_{i \in \mathcal I})$ on a phase space $(\Gamma,\{.,.\})$ we define a {\bf dual} gauge theory as a gauge theory $(H_d,\{\rho_j\}_{j\in \mathcal J})$ on $(\Gamma,\{.,.\})$ if and only if there exists a gauge fixing in the two theories such that the initial value problem and the equations of motion of both theories are identical.

One particular way to construct a dual theory is to use gauge-fixing conditions $\{\phi_i\}_{i\in\mathcal I}$ defining $\mathcal G$ such that for all $i,j\in\mathcal I$ there exist phase space functions $g^k_{ij}$ satisfying the integrability condition
\begin{equation}
 \{\phi_i,\phi_j\}=\sum_{k\in \mathcal I} g^k_{ij} \phi_k.
\end{equation}
The $\{\phi_i,.\}$ thus generate a group $\mathbb H$ of transformations on $\mathcal G$ that lets us identify $\mathcal G$ with a bundle $F$ over $\mathcal G/\mathbb H$ using an isomorphism $j:F\to \mathcal G$. The fibres of $F$ at $x\in \mathcal G$ are $\mathbb H/Iso(x)$. The Hamiltonian $H_d$ of the dual theory has to satisfy
\begin{equation}\label{equ:H-dual-conditions}
 \begin{array}{rcl}
   \left.\{H_d,.\}\right|_{\sigma}&=&\left.\{H_{gf},.\}\right|_{\sigma}\\
   \left.\{H_d,\phi_j\}\right|_{\mathcal G}&=&0.
 \end{array}
\end{equation}
These conditions as well as the integrability condition can be fulfilled by construction using the canonical St\"uckelberg formalism to implement symmetry under an Abelian group $\mathbb H$ with the subsequent elimination of the St\"uckelberg field by substituting it with the solution to the constraint equations in the original theory. The nontrivial conditions for the St\"uckelberg formalism to yield the anticipated trading of gauge symmetries are: 1) that the generators of the group furnish a gauge--fixing for the original gauge symmetry and 2) that the transformed constraint equations admit a solution in terms of the St\"uckelberg field. In the particle model, the first condition is satisfied by the invertibility of (\ref{eq: lf particle}) in terms of $N_0$ while the second condition is guaranteed by the invertibility of $T\chi_i(\phi_0) = 0$ in terms of $\phi_0$.

\section{Gravity as a Conformally Invariant Theory}

We now apply the general procedure to geometrodynamics. Before applying the procedure, we establish notation and review canonical general relativity. We treat in detail the spatially compact case leaving the asymptotically flat case for future investigations. For more details on the mathematical structure of the dualization and a more detailed proof of the dictionary, see Appendix~(\ref{appendix}).

We start with the ADM formulation of general relativity on a compact spatial manifold $\Sigma$ without boundary (and if confusion arises we will assume the topology of $\Sigma$ to be $S^3$). The phase space $\Gamma$ is coordinized by 3--metrics $g$, represented locally by a symmetric 2--tensor $g:x\mapsto g_{ab}(x)dx^adx^b$, and its conjugate momentum density $\pi$, represented locally by a symmetric 2--cotensor $\pi:x\mapsto \pi^{ab}(x)\partial_a\partial_b$ of density weight 1. Given a symmetric 2--cotensor density, $F$, and a symmetric 2--tensor $f$ we denote the smearing by
\begin{equation}
 F(g):=\int_\Sigma d^3 x F^{ab}(x) g_{ab}(x)\,\,\textrm{ and }\,\,\pi(f):=\int_\Sigma d^3x \pi^{ab}(x) f_{ab}(x).
\end{equation}
We will not explicitly state differentiability conditions for $(g,\pi)$ or details about the Banach space we use to model $\Gamma$, we just assume existence of suitable structures to sustain our construction. The nonvanishing Poisson bracket is
\begin{equation}
   \{F(g),\pi(f)\}=F(f):=\int_\Sigma d^3 x F^{ab}(x) f_{ab}(x),
\end{equation}
and the Hamiltonian is
\begin{equation}
 H(N,\xi)=\int_\Sigma d^3x\left(N(x)S(x)+\xi^a(x)H_a(x)\right),
\end{equation}
where the Lagrange mutipliers $N$ and $\xi^a$ denote the lapse and shift respectively. The constraints are
\begin{equation}
 \begin{array}{rcl}
   S(x)&=&\pi^{ab}(x)\frac{G_{abcd}(x)}{\sqrt{|g|(x)}}\pi^{cd}(x)-\sqrt{|g|(x)}R[g](x)\\
   H_a(x)&=&-2g_{ac}(x)D_b\pi^{bc}(x),
 \end{array}
\end{equation}
where $D$ denotes the covariant derivative w.r.t. $g$, $G_{abcd}$ denotes the inverse supermetric and $R[g]$ the curvature scalar. Denoting smearings as $C(f)=\int_\Sigma d^3x C(x)f(x)$ and $\vec C(\vec v):=\int_\Sigma d^3x v^a(x)C_a(x)$, we obtain Dirac's hypersurface--deformation algebra
\begin{equation}
 \begin{array}{rcl}
   \{\vec H(\vec u),\vec H(\vec v)\}&=&\vec H([\vec u,\vec v])\\
   \{\vec H(\vec v),S(f)\}&=& S(v(f))\\
   \{S(f_1),S(f_2)\}&=&\vec H(\vec N(f_1,f_2)),
 \end{array}
\end{equation}
where $[.,.]$ denotes the Lie-bracket of vector fields, so the first line simply states that the $\vec H(x)$ furnish a representation of the Lie-algebra of vector fields on $\Sigma$ and where $N^a(f_1,f_2):x \mapsto g^{ab}(x)\left(f_1(x)f_{2,b}(x)-f_{1,b}(x)f_2(x)\right)$.

\subsection{Dualization}

\subsubsection{Identify Symmetries to Exchange}

Let us spell out the symmetry to be gained in exchange for foliation invariance. Na\" ively, we would trade $S(x)$ for constraints that generate general conformal transformations of $g$. However, as in the toy model, trading all such symmetries would lead to frozen dynamics. One global constraint must be left over to generate global reparametrization invariance. In analogy to the toy model, we restrict to conformal transformations that do not change the global scale. In geometrodynamics, the analogue of the moment of inertia is the total 3--volume. Thus, the desired symmetry is explicitly constructed in the following way:

Let $\mathcal C$ denote the group of conformal transformations on $\Sigma$ and parametrize its elements by scalars $\phi:\Sigma\to \mathbb R$ acting as
\begin{equation}
 \phi: \left\{
 \begin{array}{rcl}
  g_{ab}(x)&\to&e^{4 \phi(x)}g_{ab}(x)\\
  \pi^{ab}(x)&\to&e^{-4\phi(x)}\pi^{ab}(x).
 \end{array} \right.
\end{equation}
The factor of $4$ is conventional in 3D.\footnote{In dimension $d$, the conventional factor is $\frac{4}{d-2}$.} It is chosen so that the scalar curvature $R[g]$ transforms in a simple way under the above tranformation. Consider the one--parameter subgroup $\mathcal V$ parametrized by homogeneous $\phi: x \to \alpha$. Notice that $\mathcal V$ is normal, because $\mathcal C$ is Abelian, so we can construct the quotient $\mathcal C/\mathcal V$ by building equivalence classes w.r.t. the relation
\begin{equation}
 \mathcal C \ni \phi \sim \phi^\prime \textrm{ iff }\exists \alpha \in \mathcal V, \textrm{ s.t. }\phi=\phi^\prime+\alpha.
\end{equation}
Given a metric $g$ on $\Sigma$ and $\phi \in \mathcal C$ we can find the unique representative in the equivalence class $[\phi]_\sim \in \mathcal C/\mathcal V$ that leaves $V_g=\int_\Sigma d^3x \sqrt{|g|}(x)$ invariant using the map
\begin{equation}
 \widehat{\,\,.\,\,}_g : \phi \mapsto \phi - \frac 1 6 \ln\langle e^{6\phi}\rangle_g,
\end{equation}
where we define the mean $\langle f \rangle_g:=\frac 1 V_g \int_\Sigma d^3x \sqrt{|g|}(x) f(x)$ for a scalar $f:\Sigma \to \mathbb R$. The map $\widehat{\,\,.\,\,}_g$ allows us to  parametrize $\mathcal C/\mathcal V$ by scalars $\phi$. Note that $\hat\phi_g$ can be written more transparently by observing that it is chosen so that the volume element of the conformally transformed metric is equal to
\begin{equation}
    \sqrt{ \lf| e^{4\hat\phi} g \rt| } = \frac{e^{6\hat\phi}}{\mean{e^{6\hat\phi}}} \sqrt{|g|}.
\end{equation}
Thus, the conformal transformation
\begin{equation}\label{eq conf symmetry geo}
    g_{ab}(x) \to \exp\lf( 4\hat\phi(x) \rt) g_{ab}(x)
\end{equation}
leaves
\begin{equation}
    V_g = \int d^3 x\sqrt{|g|} = \int d^3 x \frac{e^{6\hat\phi}}{\mean{e^{6\hat\phi}}} \sqrt{|g|} = V_g \frac{\mean{e^{6\hat\phi}}}{\mean{e^{6\hat\phi}}}
\end{equation}
invariant (we will often suppress the subscirpt $g$ in $\hat\phi_g$ for convenience). $\mathcal C/\mathcal V$ is then precisely the symmetry we want to obtain in exchange for foliation invariance.

\subsubsection{Introduce St\" uckelberg Field}

The next step is to perform a canonical transformation that artificially introduces the symmetry $\mathcal C/\mathcal V$ by inserting a St\" uckelberg field $\phi$. If we define the generating function
\begin{equation}
 F[\phi]:=\int d^3x g_{ab}(x) \exp\left(4\hat \phi(x)\right)\Pi^{ab}(x),
\end{equation}
then the canonical transformation $t$ generated by $F[\phi]$ is:
\begin{equation}\label{eq can trans geo}
 \begin{array}{rcccl}
   g_{ab}(x)&\to& tg_{ab}(x)&=& \exp\left(4 \hat \phi(x)\right) g_{ab}(x)\\
   \pi^{ab}(x)&\to& t\pi^{ab}(x) &=& e^{-4 \hat \phi(x)} \left(\pi^{ab}(x)-\frac 1 3 \langle\pi\rangle_g \left(1-e^{6\hat \phi(x)}\right)g^{ab}(x)\sqrt{|g|}(x)\right).
 \end{array}
\end{equation}
This is indeed a volume preserving conformal transformation.

We now enlarge the phase space $\Gamma\to \Gamma_\text{e}=\Gamma \times T^*(\mathcal C)$ with the canonical `product Poisson bracket' and parametrize conformal transformations by scalar functions $\phi$ and their conjugate momentum densities by scalar densities $\pi_\phi$. As in the particle case, to extend functions from $\Gamma$ to $\Gamma_e$ we define $T$ as $(p\circ t)^*$, where $p$ denotes the  canonical projection $p:\Gamma \times T^*(\mathcal C)=\Gamma_\text{e}\to \Gamma$. Note that $p^*$ is also a canonical transformation in the sense that the Poisson bracket of its image in $\Gamma_\text{e}$ coincides with the Poisson brackets in $\Gamma$.

The scalar constraints transform in the following way under the map $T$:
\begin{multline}\label{eq LY 1}
    S(x)\to TS(x)= \frac{e^{-6\hat\phi}}{\sqrt{|g|}} \left[ \pi^{ab} \pi_{ab} - \frac{\pi}{2} - \frac{\langle\pi\rangle^2_g}{6} (1 - e^{6\hat\phi})^2|g|+ \frac{\langle\pi\rangle_g}{3}\pi(1 - e^{6\hat\phi})\sqrt{|g|} \right]
		 \\ - e^{2\hat\phi}\sqrt{|g|} \left[ R[g] - 8\left(D^2\hat\phi + (D\hat\phi)^2 \right) \right]
 \end{multline}
where $v^2 \equiv g_{ab} v^a v^b$ for any vector $v^a$. To avoid technical difficulties arising from the proper treatment of the diffeomorphism constraint, we relax it for the time being and verify after the dualization is completed that it can be consistently reintroduced into the theory.

The quantity
\begin{equation}\label{eq D def}
    \mathcal D \equiv \pi(x)-\sqrt{|g|}(x)\langle \pi\rangle_g
\end{equation}
will be the first class constraint left over in the dual theory. We pause for a moment to note its important properties, which can be verified by straightforward calculations. First, $\mathcal D$ is invariant under the canonical transformation (\ref{eq can trans geo}). Second, it generates infinitesimal volume preserving conformal transformations. This can be seen by noting that
\begin{align}\label{eq local vpct}
    \delta_\theta\, g_{ab}(x) &= \lf( 4\theta(x) - \mean{4\theta} \rt) g_{ab}(x) \notag \\
    \delta_\theta\, \pi^{ab}(x) &= \lf( - 4\theta(x) + \mean{4\theta} \rt) \lf(\pi^{ab}(x) - \frac 1 2 \mean{\pi}_g g^{ab}(x) \sqrt{|g|}(x) \rt),
\end{align}
where $\delta_\theta  = \pb{\cdot}{\mathcal D(4\theta)}$, is the infinitesimal form of (\ref{eq can trans geo}). This is the key property that we require of the dual theory.

For the moment, the relevance of the above observation is that it allows us to show that the canonical transformation has introduced a new constraint. From (\ref{eq local vpct}), it is simple to verify that the action of $\mathcal D$ on arbitrary smooth functions $f$ of $\Gamma$ transformed by $T$ is equivalent to the action of $\pi_\phi$
\begin{equation}\label{eq vpct generator}
 \{T f,\pi_\phi(x)\}=\{Tf, \pi(x)-\sqrt{|g|}(x)\langle \pi \rangle_g\}.
\end{equation}
Thus, the quantity
\begin{equation}
 C(x)\equiv 4\left(\pi(x)-\sqrt{|g|}(x)\langle \pi\rangle_g\right)-\pi_\phi(x)
\end{equation}
acts trivially on the image of $T$. In order for this to be consistent with the dynamics, $C(x)$ should be added to the Hamiltonian as a constraint using a Lagrange multiplier. This is analogous to the toy model: the canonical transformation has both introduced auxiliary quantities $\phi(x)$ and $\pi_\phi(x)$ and corresponding constraints $C(x)$. Because of (\ref{eq vpct generator}), $C(x)$ is trivially first class with respect to any transformed quantities. Thus,
\begin{align}
    \pb{C(f_1)}{TS(f_2)} &= 0 \\
    \pb{C(f)}{T\vec H(\vec v)} &= 0
\end{align}
for the smearing functions $f_1,f_2,f$, and $\vec v$. Note that $C(x)$ could have been derived in a more conventional way as a primary constraint by introducing the St\" uckelberg field \emph{before} performing the Legendre transform. This equivalent approach was followed in \cite{Barbou:CS_plus_V}.

We note one final property of $C(x)$. Just as in the toy model, one of the $C(x)$'s is an identity and not an independent constraint. This can be seen by noting that
\begin{equation}
    \int_\Sigma d^3x\, C(x) = \int_\Sigma d^3x\, \pi_\phi(x)
\end{equation}
but the Poisson bracket of $\int d^3x\, \pi_\phi$ with the variables $g_{ij}$, $\pi^{ij}$, $\hat\phi$, and $\pi_\phi$ is identically zero (the only non--trivial calculation is $\pb{\rho(\hat\phi)}{\int_\Sigma d^3x\, \pi_\phi(x)}=0$ for arbitrary smearings $\rho$). This is a consequence of restricting to $\mathcal C/\mathcal V$ instead of just $\mathcal C$. It means that $\int d^3x\,C$ can be removed from the theory without affecting the theory in any way. The result of this is that we will be left with a global first class scalar constraint at the end of the procedure.

\subsubsection{Impose and Propagate Best--Matching Constraints}\label{sec: prop pi_phi}

We now carry out a gauge fixing inspired by best--matching \cite{Barbou:CS_plus_V,barbour:scale_inv_particles,sg:dirac_algebra}. This involves imposing the constraints
\begin{equation}
 \pi_\phi=0.
\end{equation}
One of these constraints: $\int d^3x\, \pi_\phi$, is trivial as noted above. However, singling out this constraint explicitly is more subtle now than it was in the finite dimensional toy model. For simplicity, we will work with this redundant parametrization of the constraints, keeping in mind that it is over--complete by one equation.

Imposing $\pi_\phi(x)=0$ turns all but one of the original scalar constraints into second class constraints, as can be observed from the Poisson bracket
\begin{equation}
 \{TS(N),\pi_\phi(\lambda)\}= \int dx\, \lambda(x) \left[ F_N - \sqrt{|g|}(x) e^{6\hat\phi(x)} \langle F_N \rangle_g \right]
\end{equation}
where we smeared $\pi_\phi$ with a scalar $\lambda$ and $F_N$ is given by
\begin{multline}
    F_N = 8 g_{ab} D^a\left(e^{2\hat\phi}D^bN \right)\sqrt{|g|} - 8N e^{2\hat\phi}\sqrt{|g|} \left[ R[g] - 8\left(D^2\hat\phi + (D\hat\phi)^2 \right) \right]  \\
		 - 2Ne^{6\hat\phi} \sqrt{|g|} \langle \pi \rangle_g^2 - N \left[ 6 TS + 2\pi \langle \pi \rangle_g(C + \pi_\phi) \right].
\end{multline}
The last term in $F_N$ is weakly zero. The consistency of the dynamics requires that we fix the Lagrange multipliers to satisfy the equation
\begin{equation}\label{equ:consistency-condition}
 \begin{array}{rcl}
   \{TS(N_0),\pi_\phi(x)\}&=&0.\\
 \end{array}
\end{equation}
The trivial solution $N_0\equiv 0$ yields ``frozen dynamics.'' Fortunately, there is a unique solution $N_0(\hat\phi,g_{ij},\pi^{ij})$ to the lapse fixing equation
\begin{equation}\label{equ:lapse-fixing}
    N \left[e^{-4\hat\phi} \left[ R[g] - 8\left(D^2\hat\phi + (D\hat\phi)^2 \right) \right] + \frac 1 4 \langle \pi \rangle_g^2 \right] -  e^{-6\hat\phi} g_{ab} D^a\left(e^{2\hat\phi}D^bN \right) = \mean{\mathcal L}_g
\end{equation}
that is equivalent to the first line of (\ref{equ:consistency-condition}). In (\ref{equ:lapse-fixing}), $\mathcal L$ is defined as $e^{6\hat\phi} \sqrt{|g|}$ times the left--hand--side of (\ref{equ:lapse-fixing}).

In terms of the transformed variables $G_{ab} \equiv Tg_{ab}$ and $\Pi^{ab} \equiv T\pi^{ab}$, (\ref{equ:lapse-fixing}) takes the simple form
\begin{equation}\label{eq lfe simple}
    \lf( R[G] + \frac{\mean{\Pi}_G}{4} - D^2  \rt) N = \mean{\lf( R[G] + \frac{\mean{\Pi}_G}{4} - D^2  \rt) N}_G.
\end{equation}
This is the same lapse fixing equation obtained in \cite{Barbou:CS_plus_V} by a similar argument using the Lagrangian formalism. To understand why (\ref{eq lfe simple}) has a one parameter family of solutions (and, thus, depends only on $\hat\phi$) note that any solution, $N_0$, to
\begin{equation}\label{eq lfe standard}
    \lf( R[G] + \frac{\mean{\Pi}_G}{4} - D^2  \rt) N_0 = - c^2,
\end{equation}
for an arbitrary constant $c$, is also a solution to (\ref{eq lfe simple}). Eq (\ref{eq lfe standard}) is the well--known lapse fixing equation for CMC foliations. It is known to have unique positive solutions\cite{brill:lfe_uniqueness,York:york_method_long,Niall_73}. This fact is vital for our approach to work as it allows us to construct the Hamiltonian of the dual theory.

The one parameter freedom in $N_0$ is exactly what we expect from the redundancy of the $\pi_\phi(x)$'s. We are one global equation short of completely fixing the $N(x)$'s. Thus, we should be left with a single first class linear combination of the $TS(x)$'s. This is precisely the gauge fixed Hamiltonian
\begin{equation}H_\text{gf}=\int d^3x\, N_0(x) TS(x)=TS(N_0),\end{equation}
where both $N_0(x)$ and $TS(x)$ are functionals of $g,\pi,\phi$. $H_\text{gf}$ is constructed to be first class with respect to $\pi_\phi$. We can now state the important result that \emph{the extended system with the constraint $\pi_\phi=0$ is consistent as merely a (partial) gauge fixing of the original system.}

\subsubsection{Eliminate Second Class Constraints}

Let us collect and classify the constraints of our theory to recap what we have done. We started with the first class constraints $TS(x)$ and then imposed the conditions $\pi_\phi(x) = 0$. The former are second class with respect to the latter but, since $\int d^3x \pi_\phi$ plays no role in the theory, there is a corresponding constraint $H_\text{gf}$ that is still first class. We can split the first class $H_\text{gf}$ from the remaining second class $TS(x)$'s by defining the variable $\widetilde{TS}(x) \equiv TS(x) - H_\text{gf}$. This leaves us with the first class constraints
\begin{align}
    H_\text{gf} &\approx 0 & C(x) \approx 0
\end{align}
as well as the second class constraints
\begin{align}
    \widetilde{TS}(x) &\approx 0 & \pi_\phi(x) \approx 0
\end{align}
in analogy with equations (\ref{eq 1st class part}) and (\ref{eq 2nd class part}) of the toy model. Note that, for the moment, we are still relaxing the diffeomorphism constraints. In Sec~(\ref{sec: prop pi_phi}), we showed that $N_0$ is consistent gauge choice for general relativity. However, we should now construct the Dirac bracket to explore the full structure of the theory.

First, notice that since $\{H_{gf},\pi_\phi\}\approx 0$ we know that $\{\widetilde{TS}(x),\pi_\phi(y)\} \approx \{{TS}(x),\pi_\phi(y)\}$. This leads to the weak equality
$$ \int G(x,x')\{\widetilde{TS}(x'),\pi_\phi(y)\}d^3x'\approx\int G(x,x')\{{TS}(x'),\pi_\phi(y)\}d^3x'=\delta(x,y)$$
provided the Green's function, $G$, for the differential operator acting on $N$ in (\ref{eq lfe simple}) exists. This is guaranteed because the existence and uniqueness of solutions to the CMC lapse fixing equation, for suitable initial data and boundary conditions, implies existence of the Green's function for the respective initial data and boundary value problem.\footnote{This is implied by the existence of a Green function for the so called Lichnerowicz Laplacian $D^2 +R$, an operator that can be put into the form of a Hodge Laplacian $d\delta+\delta d$.} The formal existence of this Green's function is all that we will need for the remainder of the paper.

In terms of this Green's function, the Dirac bracket is defined as:
\begin{equation}\label{Dirac geometro}
 \begin{array}{rcr}
   \{f_1,f_2\}_D&:=&\{f_1,f_2\}-\int d^3 x d ^3 y\, \{f_1,\pi_\phi(x)\}G(x,y)\{\widetilde{TS}(y),f_2\}\\
    &&+\int d^3 x d^3 y \, \{f_1,\widetilde{TS}(x)\}G(x,y)\{\pi_\phi(y),f_2\},
  \end{array}
\end{equation}
for arbitrary functions $f_1$ and $f_2$. By construction, the Dirac bracket between $\pi_\phi$ or $\wts$ and any phase space function is strongly equal to zero. We can then eliminate these constraints from the theory by following Dirac's algorithm \cite{Dirac:CMC_fixing}. First, we set $\pi_\phi = 0$ everywhere that it appears; then we find $\hat\phi_0$ such that $\wts(\hat\phi_0) = 0$ and insert $\hat\phi_0$ into all constraints. Note that it is $\hat\phi$, and not $\phi$, that we can explicitly solve for. This is analogous to what happens in the toy model and arises because of the redundancy in the $\pi_\phi(x)$. When this is done, the system will have been reduced to the original phase space $\Gamma$. The Dirac bracket between functions $f$ of $\Gamma$ and the remaining first class constraints is weakly equivalent to the Poisson bracket. This is true, just as in the toy model, because the extra terms in the Dirac bracket are either proportional to $\{f,\pi_\phi(x)\}$, which is zero, or $\{\pi_\phi(y),H_\text{gf}\}$ and $\{\pi_\phi(y),\pi_\phi(x)\}$, which are both weakly zero. Finally, because we are simply inputting the solution of a constraint back into the Hamiltonian, the equations of motion on the contraint surface will remain unchanged. Thus, $T_{\phi_0}$ is still effectively a canonical transformation. For more details, see Appendix~(\ref{appendix}) and, in particular, the steps leading to Eq~(\ref{reduced Pb}).

The last step is to verify that $\wts(\hat\phi_0) = 0$ can be solved for $\hat\phi_0$. We can simplify (\ref{eq LY 1}) using the strong equations $\pi_\phi = 0$ and by taking linear combinations of the first class constraints $C(x)$. This leads to the equivalent constraint
\begin{equation}\label{eq LY 2}
    \frac{e^{-6\hat\phi}}{\sqrt{|g|}} \lf[ \sigma^{ab} \sigma_{ab} - \frac{\mean{\pi}^2_g}{6} e^{12\hat\phi}|g| - e^{8\hat\phi} \bar R[g] \sqrt{|g|}\rt] \approx 0,
\end{equation}
where $\sigma^{ab} \equiv \pi^{ab} - \frac 1 3 \mean{\pi}_g g^{ab} \sqrt{|g|}$ and $\bar R[g] = R[g] - 8(D^2 \hat\phi + (D\hat\phi)^2)$.
Eq (\ref{eq LY 2}) is the Lichnerowicz--York (LY) equation used for solving the initial value problem of general relativity. Its existence and uniqueness properties have been extensively studied.\footnote{See \cite{Niall_73} or, for the specific context given here, see \cite{OMurchadha:LY_cspv}.} It is known to have unique solutions when $\sigma^{ab}$ is transverse and traceless. Fortunately, these are exactly the conditions required by the diffeomorphism and conformal constraints respectively. The formal invertibility of this equation is the second vital requirement for our procedure. Without this, we would not be able to prove the existence of the dual theory. However, given that we can solve (\ref{eq LY 2}) for $\hat\phi_0$ for specified boundary and initial data, we arrive at the dual Hamiltonian
\begin{equation}
    H_\text{dual}'= \mathcal N H_{gf}[\phi_0] + \int_\Sigma d^3 x \lambda(x) \left(\pi(x)-\langle \pi\rangle\right),
\end{equation}
where $\mathcal N$ is a spatially constant Lagrange multiplier representing the remaining global lapse of the theory. We can now reinsert the diffeomorphism constraint. This gives the final Hamiltonian
\begin{equation}\label{eq dual ham geo}
    H_\text{dual}= \mathcal N H_{gf}[\phi_0] + \int_\Sigma d^3 x \left[\lambda(x)\mathcal D(x) +\xi^a(x)T_{\hat\phi_0} H_a(x)\right],
\end{equation}
using the definition (\ref{eq D def}). Note that we must use the transformed $H_a(x)$ evaluated at $\hat\phi_0$ (recall that $\mathcal D$ is invariant under $T$).  As shown above, the usual Poisson bracket over $\Gamma$ can be used to determine the evolution of the system.

\subsubsection{Construct Explicit Dictionary}

We can verify that the dual Hamiltonian has the required properties. First, we check that all the constraints are first class. The gauge fixed Hamiltonian was constructed to be first class with respect to the conformal constraints $\mathcal D$. This can be seen by observing that the $TS(x)$'s are first class with respect to the $C(x)$'s and that the $C(x)$'s are equal to the $\mathcal D(x)$'s when $\pi_\phi(x) = 0$. The $H_a$'s are easily seen to be first class with $H_\text{gf}$ because they are first class with respect to the original $S(x)$'s and because $T$ is a canonical transformation. The $\pi_\phi$'s commute with themselves because they are ultralocal canonical variables. Lastly, one can directly verify that
\begin{equation}
    \pb{T_{\hat\phi_0} \vec H(\vec v)}{\mathcal D(f)} = T_{\hat\phi_0}\pb{\vec H(\vec v)}{\mathcal D(f)} = D(\mathcal L_v f) \approx 0,
\end{equation}
where $\vec v$ and $f$ are smearings.


Secondly, the dual theory is indeed invariant under volume preserving conformal transformations. For this, recall that the $\mathcal D$'s generate the infinitesimal form of (\ref{eq can trans geo}) according to (\ref{eq local vpct}). Furthermore, the theory is also invariant under 3D diffeomorphisms generated by the $H_a$'s. The theory is \emph{not}, however, invariant under 4D diffeomorphisms. The diffeomorphism invariance is only within the spatial hypersurfaces and is, thus, foliation preserving. This means, in particular, that the theory is not Lorentz invariant because it is not invariant under boosts.

Finally, there is a gauge in which the equations of motion of the two theories are equivalent. Compare the ADM Hamiltonian to that of the dual theory.
\begin{align}
    H_\text{ADM} &=\int_\Sigma d^3x\left(N(x)S(x)+\xi^a(x)H_a(x)\right) \notag \\
    H_\text{dual} &= \mathcal N H_{gf}[\phi_0] + \int_\Sigma d^3 x \left[\lambda(x)\mathcal D(x) +\xi^a(x)T_{\hat\phi_0} H_a(x)\right] \notag \\
             &= \int_\Sigma d^3 x\,\lf( \mathcal N N_0(x,\hat\phi_0) T_{\hat\phi_0} S(x) + \lambda(x)\mathcal D(x) +\xi^a(x)T_{\hat\phi_0} H_a(x)\right).
\end{align}
Because $T$ is a canonical transformation, the equations of motion of both theories take the same form in the gauges
\begin{align}\label{eq dic geo}
    N(x) &= \mathcal N N_0(x,\hat\phi_0) \notag \\
    \lambda(x) &= 0.
\end{align}
To map the solutions from one side of the duality to the other, we simply need to use the explicit function $\hat\phi_0$ and the map $T_{\hat\phi_0}$. This completes the dictionary.

This dictionary is particularly straightforward to use if one can find initial data for the ADM Hamiltonian that satisfies the initial value constraints. In this case $\hat\phi_0 = 0$. Because both theories are first class, this condition will be propagated and the solutions of each theory are \emph{equal} in the gauges (\ref{eq dic geo}). To find suitable initial data, we still must solve the LY equation. However, for the purpose of using the dictionary in the classical theory, we only need to solve this for a single point on phase space.\footnote{To prove that the dual theory actually exists and to study the quantum theory, we still must be able to solve the LY equation over all of phase space.}

Given the existence of the above dictionary, we arrive at the following proposition:
\begin{proposition}
  The theory with total Hamiltonian (\ref{eq dual ham geo}) is a gauge theory of foliation preserving 3--diffeomorphisms and volume preserving 3D conformal transformations. In the gauge $\lambda = 0$, this dynamical system has the same trajectories as general relativity in CMC gauge.
\end{proposition}

\section{Conclusions and Outlook}

Let us summarize the main result of this paper: we showed that there is a duality between general relativity and a gauge theory of 3D conformal diffeomorphisms that preserve the total spatial volume. Such a theory does not admit a local scale and is, therefore, a theory of the dynamics of the intrinsic ``shape'' of objects. We identify this theory with Barbour and \'O Murchadha's \emph{shape dynamics}. The classical duality means that there exists a distinguished gauge in both theories such that the gauged trajectories of one theory coincide precisely with the gauged trajectories of the other. In other words, one can trade the gauge symmetry of general relativity (spatial diffeomorphisms and local refoliations) for the gauge symmetry of shape dynamics (spatial diffeomorphisms, volume preserving conformal transformations, and global refoliations) without changing the physical content of the theory. The Hamiltonian of shape dynamics is nonlocal and difficult to write explicitly. However, explicit expresssions can be obtained, for example, by applying our procedure to the torus universe in 2+1 dimensions. One can draw a number of physical conclusions from this duality:
\begin{enumerate}
 \item Physical degrees of freedom of general relativity: The physical degrees of freedom of general relativity and shape dynamics are identical. This confirms earlier results \cite{York:york_method_prl,Barbou:CS_plus_V,Barbour:shape_dynamics} that the physical degrees of freedom of general relativity can be labeled by conformal 3--geometries and their momenta.
 \item Theory space of quantum gravity: The only input into the asymptotic safety approach to quantum gravity is the theory space. While the theory space of general relativity is usually identified with 4--diffeomorphism invariant metric theories, the theory space of shape dynamics is 3D conformal diffeomorphism invariant metric theories. This difference in theory space may lead to a different fixed point structure and different resolutions to the ghost problem of quantum gravity. Furthermore, the theory space of shape dynamics is locally identical to that of Ho\v rava--Lifshitz gravity. This provides a possible link between Ho\v rava--Lifshitz gravity and shape dynamics.
 \item Loop Quantum Shape Dynamics: Following the Loop Quantization program, one can quantize shape dynamics. Observing that scale information in Loop Quantum Gravity is encoded in the spin quantum numbers on the edges and the intertwiner quantum numbers on the vertices of spin network functions, one may speculate that the physical Hilbert space of Loop Quantum Shape Dynamics is spanned by unlabeled graphs carrying a total volume quantum number. With this proposal, one should study Hamiltonians on this physical Hilbert space that propagate a CMC condition.
 \item New possibilities for quantum geometrodynamics: Since all constraints of shape dynamics generate transformations that preserve the configuration space of all 3--metrics, one can implement the gauge transformations as pullbacks on this configuration space. The corresponding operator algebra is naturally represented on a kinematic Hilbert space in which each distinct 3--metric defines a normalized geometry eigenstate orthogonal to all other geometry eigenstates. The implementation of this is currently a work in progress by the authors.
\end{enumerate}
Lastly, the duality between general relativity and shape dynamics is not accidental. There is a general mechanism at work that can be seen in a variety of classical gauge theories. This is a consequence of the second result of this paper: given a gauge fixing that satisfies certain inegrability conditions, one can construct a dual theory by replacing the original gauge generators with the gauge fixing conditions.

\section{Acknowledgments}

We would like to thank Julian Barbour and Lee Smolin for helpful discussions and for providing useful comments on the draft. We also like to thank Niayesh Afshordi for his help and interest. HG would like to thank John Barrett for support. This project was born out of stimulating visits to College Farm but was raised on the many chalk boards of the Perimeter Institute. We are indebted to these rich environments for creating atmospheres that promote independent thought. Research at the Perimeter Institute is supported in part by the Government of Canada through NSERC and by the Province of Ontario through MEDT.

\appendix

\section{Direct phase space geometry of exchange}\label{appendix}

We now give a direct and formal proof of the steps involved in the dualization in geometrodynamics.

To justify the use of geometrical arguments, we should have that the regular value set
$$\bigcap_i\{(\chi_i)^{-1}(0)\},$$
 for our constraints $\chi_i$, should initially form a first class submanifold of phase space. In the infinite--dimensional context, the immersions need to be splitting. This is a non--trivial condition that does not follow solely from the finite--dimensional result of manifolds formed from regular points. For definiteness, let us take
 the constraints to be $\mathcal{H}$ and $\mathcal{H}^a$. For these, one can (almost\footnote{For the proof that this intersection is indeed a manifold, one needs to further use the constraint  that $\pi=c(t)$, i.e. that the mean extrinsic curvature be a spatial constant. This is precisely the condition that arises in our procedure.}) show that the intersection of the regular points is indeed a manifold  using the Fredholm alternative \cite{Fisher_and_Marsden,Hawking:Einstein_centenary}. Now we are ready to state our process in a geometrical setting.

 \subsection{Formal Steps}
 \begin{enumerate}
 \item {\bf Extend ambient phase space}

 We define the extended ambient manifold, $\Gamma_{\mbox\tiny E}:=\Gamma\times T^*\mathcal{C}$, by using a global parametrization of the additional coordinates through $(\phi, \pi_\phi)$.  Restricting to the subspace parametrized by $\hat\phi$ (which is nonetheless more conveniently parametrized, albeit redundantly, by $\phi$) we find that $\langle\pi_\phi\rangle$ has trivial action on all of the variables. Thus, we can formally restrict our attention to the space
     \begin{equation}\Gamma_e:=\Gamma\times{T^*(\mathcal{C}/\mathcal{V})}\end{equation}
     conveniently (and redundantly) parametrized by $(g_{ij},\pi^{ij},\phi,\pi_\phi)$. Note that the redundancy has a nontrivial dependence on $\Gamma$.

    \item {\bf St\"uckelberg Mechanism}

     For $\mathcal{F}:\Gamma\rightarrow C^{\infty}(\Sigma)$, we define the natural extension
\begin{equation}T{\mathcal{F}}:\Gamma_e\rightarrow C^{\infty}(\Sigma),\end{equation} This is the simple transformation $T\mathcal{F}(g_{ij},\pi^{ij})=\mathcal{F}(T_\phi g_{ij},T_\phi \pi^{ij})$. Under this extension, we have the following identities:
\begin{equation}\{\mathcal{D}(\theta),T g_{ab}\} = \{\pi_\phi(\theta),Tg_{ab}\}~~\mbox{and}~~\{\mathcal{D}(\theta), T\pi^{ab}\} =\{\pi_\phi(\theta),T\pi^{ab}\},
\end{equation}
 where $\theta(x)$ is a smearing function. Thus, we have the following quantity acting trivially on the image of $T$:
\begin{equation} C(x)\equiv \pi(x)-\sqrt{g}(x)\langle \pi\rangle_g-\pi_\phi(x).
\end{equation} Imposing this constraint on functions on $\Gamma_e$ is an efficient way to require our dynamical systems to be in a one to one relation with the one defined in the original $\Gamma$. This constraint, arising from the St\"uckelberg mechanism, eliminates the extra degrees of freedom that have been introduced. Because of our redundant parametrization, $C(x)$ contains the redundant constraint $\langle C\rangle=0$.

\item {\bf Impose $\mathbf{\pi_\phi=0}$ and separate first and second class parts.}

 $\pi_\phi = 0$ trivially forms a first class submanifold of $\Gamma_e$. Its symplectic flow is equivalent to that of $\mathcal{D}$ on the image of $T$ and it is thus tangent the two first class submanifolds $C(x)\approx 0$ and $T_\phi\mathcal{H}^a\approx 0$:
 \begin{equation} \{C,\pi_\phi\}=0~~\mbox{and}~~
\{T\mathcal{H}^a(N_a),\pi_\phi\}\approx0.
 \end{equation}Their intersection consists of the lift through the $T$ map of the space of transverse momenta with spatially constant trace given by $\langle\pi\rangle$. This can be shown to form a manifold.

 Regarding the space generated by the remaining constraints $T\mathcal{H}(x)$ (one at each point), there is still one direction, among the infinitely many, that is tangent to \begin{equation}\bigcap_{x\in\Sigma}\{\pi_\phi(x)\}^{-1}(0)\bigcap_{y\in\Sigma} \{T\mathcal{H}(y)\}^{-1}(0).\end{equation}
 This is
 \begin{equation} \{TS(N_0),\pi_\phi(x)\}\propto TS(x)\approx0 \label{equation above}
 \end{equation}
 where $N_0[g_{ij};\phi,x)$. Thus, the direction given by $H_\text{gf}=\int d^3x\, N_0(x) TS(x)$  is first class. Furthermore, (\ref{equation above}) has a homogeneous term that includes the integral of $N$ so that its solution is only determined up to a constant scaling: i.e. if $N_0$ is a solution so is $aN_0$ for $a\in\mathbb R/\{0\}$. We, therefore, have existence and uniqueness of solutions only on the domain $C^\infty(\Sigma)/\mathcal{V}$.
 We can split $TS(x)$ into a part that is first class with respect to all the constraints and a part that is second class:
 \begin{equation} TS(N_0) ~~\mbox{and}~~\widetilde{TS}(x):=TS(x)-TS(N_0),
 \end{equation}
 respectively.

  \item[$\bullet$] {\bf Intemezzo}

  We now show how these properties arise. The homogeneous lapse fixing equation resulting from the Poisson bracket $\{H(N),\pi(x)\}=0$ is
 \begin{equation}(\nabla^2+f(x))N(x):=\Delta N(x)=0,\label{homogeneous}
 \end{equation}where we define the second order elliptic self--adjoint differential operator $\Delta=(\nabla^2+f(x))$ and $f(x) = -(1+R)$ in the case we are considering. There are many different ways to prove that, as a differential operator, $\Delta$ is an isomorphism from smooth functions to smooth functions (see for example \cite{Niall_73} or \cite{Friedlander} for the construction of the respective Green's functions\footnote{This can also be extended to distributional domains.}). 
Taking this to be the case, $\Delta$ is an invertible operator to which one can find solutions for appropriate boundary data.

 The inhomogeneous analogue to (\ref{homogeneous}) is
 \begin{equation}\label{non-homogeneous}\Delta N (x)-\langle \Delta (N)\rangle=\{H(N),(\pi-\langle\pi\rangle\sqrt{g})(x)\}=0. \end{equation}
 Using the above arguments, we have a one parameter family of solutions to \eqref{non-homogeneous}, that is, whenever $\Delta N (x)-\langle \Delta (N)\rangle=c$ where $c$ is a given non-zero constant.  Suppose we have a solution, $N_0(x,g_{ij}]$, such that $c=1$. Applying the canonical transformation $T$, we get
 \begin{equation}\label{above eq 2}(T\Delta) N(x)-\langle (T\Delta) N\rangle=\{TH(N),(\pi-\langle\pi\rangle\sqrt{g})(x)\}= T\{H(N),(\pi-\langle\pi\rangle\sqrt{g})(x)\}.\end{equation}
 A solution, $\tilde N_0(x;g_{ij},\phi]$, to (\ref{above eq 2}) is
 \begin{equation} \tilde N(x;g_{ij},\phi]=T_\phi N(x,g_{ij}],
 \end{equation}where $N(x,g_{ij}]$ is the solution to the untransformed equation \eqref{non-homogeneous}.
 \item[$\bullet$]{\bf End of intermezzo}

 \item {\bf Solving the constraints}

   For the unconstrained conformal transformations, which we will denote by $\bar T$, \begin{equation}{\bar {T}S}(x):\Gamma\times T^*(\mathcal{C})\rightarrow \mathcal{C},\end{equation} where we used that $C^\infty(\Sigma)\simeq \mathcal{C}$. Since these equations do not depend on $\pi_\phi$, we can fix $\pi_\phi(x)=f(x)$. Then
  \begin{equation} {\bar {T}S}(x)_{\pi_\phi=f(x)}:\Gamma\times C^\infty(\Sigma)\rightarrow C^\infty(\Sigma).
\end{equation} Denoting the derivative in the second coordinate, the one parametrized by $\phi$, by a subscript $\mathcal{C}$, we have, from equation \eqref{homogeneous}, that the linear operator
\begin{equation} \delta_{\mathcal{C}}\bar{T}S:=\frac{\delta {\bar{T}\mathcal{H}(x)}}{\delta\phi(y)}=\{\bar{T}H(x),\pi_\phi(y)\}=\Delta(x)\delta(x-y)
\end{equation}
  is a topological linear isomorphism\footnote{We assume that, for all practical purposes, we can carry on as if these were Banach spaces.} between the spaces $C^\infty(\Sigma)$ and $C^\infty(\Sigma)$. However, for the volume preserving conformal transformations \begin{equation} {TS}(x)_{\pi_\phi=f(x)}:\Gamma\times C^\infty(\Sigma)/\mathcal{V}\rightarrow C^\infty(\Sigma),
\end{equation} we explicitly derived, from properties of the full conformal transformations, that the linear operator
\begin{equation} \delta_{\mathcal{C}}TS:=\frac{\delta {T\mathcal{H}(x)}}{\delta\phi(y)}=\{TH(x),\pi_\phi(y)\}
\end{equation}
  does not yield a topological linear isomorphism between the spaces $C^\infty(\Sigma)/\mathcal{V}$ and $C^\infty(\Sigma)$.

   The adjoint ``matrix" has a one dimensional kernel given by $TH(N_0)$. However
  $$ \widetilde{TS}(x)_{\pi_\phi=f(x)}:\Gamma\times C^\infty(\Sigma)/\mathcal{V}\rightarrow C^\infty(\Sigma)/\mathcal{V}
$$ has the infinitesimal transformation
  \begin{equation} \delta_{\mathcal{C}}\widetilde{TS}:=\frac{\delta {T\mathcal{H}(x)}}{\delta\phi(y)}=\{\widetilde{TH}(x),\pi_\phi(y)\}.
\end{equation}
Having taken out the kernel, this is necessarily an invertible operator. Not only can we form the Dirac bracket using $\{\widetilde{TH}(x),\pi_\phi(y)\}^{-1}$ but we can now use the implicit function theorem for Banach spaces to assert that
\begin{proposition}
 There exists a unique $\hat\phi_0:\Gamma\rightarrow \mathcal {C}/\mathcal{V}$ such that
 $$ (\widetilde{TS})^{-1}(0)=\{(g_{ij},\pi^{ij},\hat\phi_0[g_{ij},\pi^{ij}],\pi_\phi)~|~(g_{ij},\pi^{ij})\in\Gamma\}.
$$
\end{proposition}

 \item {\bf Construct the theory on the constraint surface}

We have then a surface in $\Gamma_e$, defined by $\pi_\phi=0$ and $\phi=\phi_0$, on which $\widetilde{T\mathcal{H}}=0$, and whose intrinsic coordinates are $g_{ij},\pi^{ij}$.
Furthermore, the Dirac bracket on the surface exists and we can use it to help prove theorems on the constraint surface, which now has the symplectic structure
\begin{equation}\label{reduced Pb}\{\cdot,\cdot\}_{|{\mbox\tiny{reduced}}}:={\{\cdot,\cdot\}^{\Gamma_e}_{\mbox{\tiny{DB}}}}_{|\phi=\phi_0,\pi_\phi=0}
    =\{\cdot_{|\phi=\phi_0,\pi_\phi=0},\cdot_{|\phi=\phi_0,\pi_\phi=0}\}.\end{equation}

    One can immediately see from \eqref{reduced Pb} that the first class constraints $\mathcal{D},T\mathcal{H}^a$ and $ \langle T N_0\mathcal{H}\rangle$ remain first class. Furthermore, we can verify that
    \begin{equation} \{\cdot,T_\phi\chi\}_{|{\mbox\tiny{reduced}}}= T_{\phi_0}\{\cdot,\chi\}\end{equation} for the remaining first class constraints $\chi$ (as they are represented in the original phase space). Thus, $T_{\phi_0}$ is a canonical transformation for any of the first class constraints, giving us all we need for the dynamics. Finally, we reconfirm the conclusions of the main text: that we have a dual Hamiltonian
    \begin{equation} H_{\text{dual}}=\mathcal N\langle N_0\mathcal{H}\rangle+\int_\Sigma d^3 x \left(\lambda(x) \left(\pi(x)-\langle {\pi}\rangle\sqrt{g}\right)+\rho^a(x)H_a(x)\right)\end{equation} in $\Gamma$ with the first class constraints
    \begin{equation}\langle N_0\mathcal{H}\rangle,~\mathcal{D},\mathcal{H}^a\end{equation} and the physical configuration space
    $\mathcal{M}/(\mbox{Diff}(\Sigma)\times (\mathcal{C}/\mathcal{V}))$.

    \item[$\bullet$] {\bf Final remark}

    We could have used the operator
    \begin{equation}
     (\bar{T}\Delta)\delta(x,y)=\{\bar {T}H(x),\pi(y)\}=\{\bar{T}H(x),\pi_\phi(y)\},
    \end{equation}
    which generates the \emph{full} group of conformal transformations, instead of $T$. This operator is invertible because of the invertibility of $\Delta$. Also, we could have both used the implicit function theorem and formed a Dirac bracket. However, the homogeneous equation only has the solution $N=0$ for compact manifolds without boundary. These don't generate any dynamics.

 \end{enumerate}

\bibliographystyle{utphys}
\bibliography{shape_dynamics}

\end{document}